\documentclass[preprint2]{aastex}
\shorttitle{Local Group and other galaxy groups}
\shortauthors{Karachentsev}

\begin{document}
\title{The Local Group and other neighboring galaxy groups\altaffilmark{1}}
\author{I. D. Karachentsev}
\affil{ Special Astrophysical Observatory of Russian Academy
               of Sciences, N.Arkhyz, KChR, 369167, Russia}
\email{ikar@luna.sao.ru}

\altaffiltext{1}{Based in part on observations made with the NASA/ESA Hubble
Space Telescope.  The Space Telescope Science Institute is operated by
the Association of Universities for Research in Astronomy, Inc. under NASA
contract NAS 5--26555.}

\begin{abstract}
  Over the last few years, rapid progress has been made in distance
measurements for nearby galaxies based on the magnitude of the tip of
red giant branch stars. Current CCD surveys with HST and large ground-
based telescopes bring  $\sim$10\%-accurate distances for roughly a hundred 
galaxies within 5 Mpc. The new data on distances to galaxies situated
in (and around) the nearest groups: the Local Group, M81 group, CenA/M83
group, IC342/Maffei group, Sculptor filament, and Canes Venatici cloud
allowed us to determine their total mass from the radius of the zero-
velocity surface, $R_0$, which separates a group as bound against
the homogeneous cosmic expansion. The values of $R_0$ for the virialized
groups turn out to be close each other, in the range of 0.9 -- 1.3 Mpc.
As a result, the total masses of the groups are close to each other,
too, yielding total mass-to-blue luminosity ratios of
10 -- 40 $M_{\sun}/L_{\sun}$.
The new total mass estimates are 3 -- 5 times lower than old virial mass
estimates of these groups. Because about half of galaxies in the Local
Volume belong to such loose groups, the revision of the amount of dark matter
(DM) leads to a low local density of matter, $\Omega_m \simeq 0.04$, which is
comparable with the global baryonic fraction $\Omega_b$, but much lower than
the global density of matter, $\Omega_m = 0.27$.

To remove the discrepancy between the global and local quantities
of $\Omega_m$, we assume the existence of two different DM components:
1) compact dark halos around individual galaxies and
2) a non-baryonic dark matter ``ocean'' with $\Omega_{dm1} \simeq
0.07$ and $\Omega_{dm2} \simeq 0.20$, respectively.

\end{abstract}
\keywords{ galaxies: distances and redshifts --- galaxies: kinematics
and dynamics --- galaxies: individual ( Milky Way, M31, M81, Cen A,
M83, IC 342, Maffei 1, NGC 253, NGC 4736 ) --- galaxies:
dark matter --- cosmology }

\section{Introduction}
  In rich clusters of galaxies, virial mass estimates agree well with
independent determinations of the cluster mass made from the X-ray flux
of hot intracluster gas, and from weak gravitational lensing effects.
A typical ratio of the total mass-to-blue luminosity for rich clusters,
$M_t/L_B \sim 250 M_{\sun}/L_{\sun}$, extrapolated over the whole
volume of the universe, yields the mean density of matter
$\Omega_m \sim$0.25, in the excellent concordance with parameters of the
standard $\Lambda$CDM model: $\Omega_m$ = 0.27, $\Omega_{\Lambda}$ = 0.73
(Spergel et al. 2003). However, about 85\% of galaxies are
situated outside the rich clusters. Roughly a half of them belong
to groups of different size and population, while the remaining half are
scattered in diffuse (unvirialized) ``clouds'' and ``filaments'' usually
called the ``field''. Until recently, application of the virial theorem
to galaxy groups remained the only way to trace the dark matter
distribution on scales of 0.1 -- 1 Mpc. Measurements of the total mass
of individual groups via their X-ray flux or weak gravitational lensing
have not lead yet to distinct results.

  In the case of our Local Group (LG), Lynden-Bell (1981) and Sandage (1986)
proposed to determine its total mass using a method, which is based on
a measurement of the radius of the ``zero-velocity surface'', $R_o$,
where the overdensity of the group has halted expansion and infall is
about to commence.  Under the assumption of spherical symmetry the
total mass of a group can be expressed in terms of the radius $R_o$,
the age of the Universe $T_o$ , and the Gravitational
constant $G$ as
   $$M_t = (\pi^2/8G)\times R_o^3 \times {T}_o^{-2}.$$
Application of this method to the LG as well to other groups requires
accurate distances and radial velocities for all galaxies surrounding
a group. Such data is now accessible for several nearby groups by using
the luminosity of stars at the tip of the red giant branch (TRGB) as
standard candles.

  Over the last 5 years, searches for additional nearby dwarf galaxies,
made on the POSS-II \& ESO/SERC plates by Karachentseva \& Karachentsev
(1998; 2000), and also results of ``blind'' HI surveys of the southern
sky, and the Zone of Avoidance by the HIPASS team
(Kilborn et al. 2002), as well as other wide-field sky surveys,
lead to a doubling of the population of known galaxies in the Local volume.
Basic observational data on 450 Local volume galaxies with distances
less than 10 Mpc are presented in the Catalog of Neighboring Galaxies
(Karachentsev et al. 2004).

 The new quantity and quality of observational data on nearby galaxy
distances and radial velocities allowed us to determine virial and total
masses of four of the nearest complexes: Milky Way+M31 (Karachentsev et al.
2002c), M81+NGC2403 (Karachentsev et al. 2002a), CenA+M83 (Karachentsev
et al. 2002b), and IC342+Maffei (Karachentsev et al. 2003d), as well
as of two other nearby scattered galaxy systems:  M96 = the Canes Venatici I
cloud (Karachentsev et al. 2003a) and NGC253 = the Sculptor filament
(Karachentsev et al. 2003c). Below we discuss the results of mass
determination for all six nearest groups made via internal and external
galaxy motions.

\section{ Some particular properties of the neighboring groups}

   Our Galaxy, the Milky Way, and the Andromeda nebulae, M31,
together with their suites of dwarf companions are usually combined
into the united dynamical system, the Local Group (=LG).
The basis for this is the observed mutual approaching of two the giant
galaxies with a velocity of 123 km s$^{-1}$. However, such a definition has an
apparent disadvantage because the dynamical center of the LG turns out to
be in emptiness, and individual motions of dwarf galaxies occur around the
Andromeda and Milky Way themselves. For this reason, we prefer to consider
both the giant spiral galaxies as the centers of two independent groups
forming together the Local complex of galaxies.

\subsection{\em The Milky Way group.}
Different properties of companions to our
Galaxy are described in detail in a monograph by van den Bergh (2000)
``The galaxies of the Local Group''. We will discuss here only the most
general parameters of the group. The list of galaxies dynamically
associated with the Milky Way is presented in Table 1. Its columns
contain: (1) galaxy name; (2) morphological type; (3,4) radial velocities
with respect to the Galaxy center and regarding to the LG centroid,
respectively, with the apex parameters adopted in the NASA Extragalactic
Database (NED); (5) absolute magnitude corrected for the Galactic
extinction (Schlegel et al. 1998) and internal galaxy extinction in the
manner, which takes into account the galaxy inclination, as well as
luminosity (Karachentsev et al. 2004); (6) distance to the galaxy in
Mpc with indication of the method used: from the luminosity of cepheids
(cep), from the luminosity of the tip of the red giant brench (rgb) as
given in (Karachentsev et al. 2004); (7) the so called ``tidal index'':
$$\Theta_i = \max  \{\log(M_k/D_{ik}^3)\} + C,\;\;\;  (i = 1, 2...  N)$$
where  $M_k$ is the total mass of any neighboring galaxy separated from
the considered galaxy by a space distance $D_{ik}$; for every galaxy ``$i$''
we found its ``main disturber''(=MD), producing the highest tidal action; the
value of the constant $C$  is chosen so that $\Theta=0$ when the Keplerian
cyclic period of the galaxy with respect to its main disturber equals the
cosmic Hubble time, $1/H$. In this sense, galaxies with $\Theta<0$ may be
considered as isolated objects of the general field. Besides 15 members of
the Milky Way group, we indicate under the horizontal line at the bottom of
Table another four galaxies: Tucana, SagDIG, SexA, and SexB, for which the
Milky Way is also the MD. All of them move away from our Galaxy,
taking apparently part in the general cosmic expansion. Note that
the LMC stands out as the MD with respect to the Milky Way and the SMC.

  By luminosity, the Milky Way appreciably dominates over the surrounding
objects: the net luminosity of its 14 companions is an order magnitude
lower than the Milky Way luminosity itself. The radius of gravitational
prevalence of the Milky Way extends to $\sim$700 kpc, while the radial
velocity dispersion of the companions is 86 km s$^{-1}$ with regard to the
Galaxy center and 76 km s$^{-1}$ with respect to the LG centroid. Applying the
virial theorem, we obtain for our group the mass estimate
$$M_{vir} = 3\pi N \times(N-1)^{-1} \times G^{-1} \times \sigma^2_v \times R_H,$$
where $\sigma_v^2$ is the dispersion of radial velocities with respect to the
group centroid, and $R_H$ is the mean projected harmonic radius.
This relation assumes a spherical symmetry of the group and a random
orientation of velocity vectors for the group members.
Adopting $R_H = (\pi/4) \langle D_i^{-1}\rangle^{-1}$ = 68 kpc, gives us  $M_{vir} =
93\times 10^{10} M_{\sun}$. For arbitrarily oriented Keplerian orbits
of companions with the eccentricity $e$  the robust estimator of mass is
$$M_{orb} = (32/3\pi)\times G^{-1}\times(1- 2e^2/3)^{-1} \langle R_p\times \Delta V^2_r
\rangle.  $$
Adopting $e$ = 0.7 as the average eccentricity, we derive for the companions
of the Milky Way  $M_{orb} = 96\times 10^{10} M_{\sun}$ in excellent agreement
with the previous estimate.

  However, two circumstances exist that force one to suppose that the two
evaluations obtained overestimate the mass of our group. Firstly,
all the companions to Milky Way (but for NGC 6822) are situated
in an elongated volume with an axial ratio 8:3:1, known as the polar
Magellanic stream. Secondly, when observing orbital motions of the
companions from inside, if orbits are strongly elongated then we see
almost the full
vectors of their velocities, i.e. the statistical relationship
$\sigma^2_{v_r} = (1/3) \sigma_v^2$ between radial and spatial velocity
dispersions is not satisfied.

\subsection{\em The Andromeda group.}
Over the last five years five new companions to M 31: And V, Cas dSph,
Peg dSph, Cetus, and And IX (Zucker et al. 2004) have been
discovered. As a result, the present population of M 31 group exceeds the
population of the group around our Galaxy. The list of 19 members of the
Andromeda group is given in Table 2. The following information is provided
in the Table: (1) galaxy name; (2) morphological type; (3) radial velocity
with respect to the LG centroid; (4) absolute magnitude after corrections
for internal and external extinctions; (5) distance from the observer;
(6) angular separation from the M 31, and (7) the tidal index. As it is
seen, the maximum angular separation of the most remote companions to
Andromeda reaches one radian. At the bottom of the table under the horizontal
line another four dwarf galaxies are given: DDO 210, KKH 98, KKR 25,
and KK 230, for which the M 31 is the MD. However, their crossing
time with respect to the Andromeda exceeds the age of the universe,
which does not permit us to consider them physical members of the M 31 group.
For 18 companions to Andromeda their mean projected linear separation is
254 kpc, while the mean harmonic radius is 42 kpc. The radial velocity
dispersion in the M 31 group, 77 km s$^{-1}$, is nearly the same as
in the Milky Way group. Applying the above mentioned relations, we
obtain the mass estimates:  $M_{vir} = 57\times 10^{10} M_{\sun}$ and $M_{orb} =
111\times 10^{10} M_{\sun}$. Studying the kinematics of the giant stellar stream
around the M 31, Ibata et al. (2004) measured the mass of Andromeda halo on
the scale of 125 kpc. Their estimate $M_{125} =(75^{+25}_{-13})\times 10^{10}
M_{\sun}$ agrees well with the virial and orbital mass estimates.
Note also, that almost the same mass estimates have been derived by
Evans \& Wilkinson (2000), Cote et al. (2000), and Evans et al. (2000)
on the scale of $\sim$50 kpc based on the motions of companions and globular
clusters of M 31.

\subsection{\em The M 81 group.} Kinematics of galaxies around the M 81
has been considered by Karachentsev et al. (2002a). Recently,
several new galaxies of the group have been imaged with the Advanced
Camera for Survey at the HST. A slightly renewed
parameters of the M 81 group members are summarized
in Table 3, where column designations  are the same as in previous
tables ( In column 6 ``mem'' means that the galaxy distance was ascribed
from its membership as the mean distance to the group.) The last column
contains notes regarding some objects, for which the main disturber is
not M 81, but other nearby galaxies. As it is seen, in the number of
members, N = 29, the M 81 group exceeds both the Milky Way group and
the Andromeda group. Below the horizontal line in the Table 3, we give
another six galaxies which may belong also to the far periphery of the
M 81 group. Some of them
are associated with another spiral galaxy NGC 2403 rather
than with M81 itself, forming a scattered system (``flock''), which moves
towards the M 81. Both from the radial velocity dispersion, $\sigma_v$ = 91
km s$^{-1}$, and from the linear projected radius, $\langle R_p\rangle = 211$ kpc,
the M 81 group is similar to the Milky Way and the M 31 groups. Virial and
orbital mass estimates for the M 81 group are $117\times 10^{10} M_{\sun}$ and
$197\times10^{10} M_{\sun}$, respectively.

\subsection{\em The CentaurusA/M83 complex.} The structure and kinematics
of this nearby galaxy complex have been discussed by Karachentsev et al.
(2002b). The group of galaxies around M 83 (=NGC 5236) actually has
the same mean radial  velocity (+308 km s$^{-1}$)
as the Cen A group (+312 km s$^{-1}$), but the distance from us appreciably
larger (4.56 Mpc) than that of the Cen A group (3.66 Mpc). The updated lists
of 28 and 14 members of both the groups are presented in Tables 4 and 5,
respectively. Notes in the last column indicate group members having another
main disturbers (not Cen A or M 83). Under the horizontal lines there
are some galaxies with larger projected separations which may be
associated with the complex too.

  In the Cen A group its mean linear projected radius, 290 kpc, and
radial velocity dispersion, 105 km s$^{-1}$, are markedly greater than in
the M 83 group ( 164 kpc and 71 km s$^{-1}$) that leads to a considerable
difference in mass estimates of the groups: 489 (vir) and 288 (orb) in
$10^{10} M_{\sun}$  for the Cen A, and 109 (vir) and 100 (orb) in $10^{10} M_{\sun}$
for the M 83 group, respectively. Such a difference is consistent with the
assumption of Bahcall et al. (1995) that giant elliptical galaxies
have a mass 2 -- 3 times larger per unit luminosity than giant spiral
galaxies.

\subsection{\em The IC342/Maffei complex.} This nearby binary group is
situated in the zone of strong Galactic extinction, which complicates analysis
of its dynamics. Properties of the structure and kinematics of the complex
have been considered by Karachentsev et al. (2003d). The last summary
of data on distances, radial velocities and luminosities of galaxies
around IC342/Maffei was published by Karachentsev et al. (2004). These
data are reproduced in Table 6 and 7. Here, distance estimates made from
luminosity of the brightest stars or via Tully-Fisher and Faber-Jackson
relations are indicated as ``bs'', ``tf'', and ``fj'', respectively. For
three galaxies: UGCA 86, KKH 37, and KKH 6 we use here new distance
estimates obtained with the ACS HST.

  At present, both the groups around the giant face-on spiral IC 342, and
around the pair of E+S galaxies Maffei~1 and Maffei~2 contain eight members each.
Their numbers may be increased after a careful survey of this region in the
HI line.  Two galaxies, KKH 37 and KKH 6, with negative tidal indexes
(in bottom of tables 6 and 7) do not belong, apparently, to the bound
members of the complex.

  Both the groups have a rather low dispersion of radial velocities,
54 km s$^{-1}$ (IC 342) and 59 km s$^{-1}$ (Maffei), and quite typical mean
linear projected separations, 322 kpc and 104 kpc, respectively. The virial
and orbital mass estimates for the groups come to 57 and 95 (IC 342),
and 65 and 135 (Maffei) in units of $10^{10} M_{\sun}$.

\subsection{\em The Sculptor filament.} As it was shown by Jerjen et al.
(1998), a conglomeration of bright galaxies in Sculptor is a loose
filament stretched along the line of sight and involved in the
general cosmic expansion. According to Karachentsev et al. (2003c),
the giant spiral galaxy NGC 253 and its five companions form a
semi-virialized core of the filament. Data on NGC 253 and its companions
are given in Table 8. In its lower part we indicate also five more
remote galaxies associated with the group core. At a radial velocity
dispersion 54 km s$^{-1}$ and a mean harmonic projected radius
$R_H = 347$ kpc, estimates of mass of the group
are 332 (vir) and 153 (orb) in units of $10^{10} M_{\sun}$.
However, both the mass estimates look extremely unreliable since the group
crossing time, $T_{cross} = \langle R_p\rangle /\sigma_v = 6.6$ Gyr, is too long
for the virialization of the system. Note also, that inclusion of 5 more
distant galaxies under the horizontal line in Table 8 brings the crossing
time closer to the Hubble time, $\sim$13 Gyr.

\subsection{\em The Canes Venatici I cloud.} This very loose extended system
mainly inhabited by dwarf irregular galaxies was studied by Karachentsev
et al. (2003a). On the map of the sky region presented by the authors
the CVnI cloud occupies an area of about 35$\degr$ in diameter
with the center near the brightest galaxy M 94 (=NGC 4736). For the sample
of 34 members of the cloud, Karachentsev et al. (2003a) derived the radial
velocity dispersion  50 km s$^{-1}$, the mean linear projected radius
760 kpc, and the mass estimates: $360\times 10^{10} M_{\sun}$ (vir) and
$190\times 10^{10} M_{\sun}$ (orb).The crossing time for the CVnI cloud
is 15 Gyr, therefore, the system is rather in the free Hubble expansion
than in a state of dynamical equilibrium.

  Arrangement of galaxies by their tidal index shows that
the brightest galaxy of the CVn I cloud, M 94, is the main disturber
relative only to 10 neighboring galaxies. All of them are presented in
Table 9, where column designations are the same as in the previous tables.
For 9 members of the M 94 group with positive tidal indices, we obtain
the radial velocity dispersion 56 km s$^{-1}$, the mean harmonic projected
radius 346 kpc, the mass estimates $267\times 10^{10} M_{\sun}$ (vir),
$322\times 10^{10} M_{\sun}$ (orb), and the crossing time 6.9 Gyr.
Thus, even the central region of the CVnI cloud can not be
considered as dynamically relaxed sub-system.

  \section{ ``Bald'' dwarfs in the groups}
  In all the groups, but for the strongly obscured complex IC342/Maffei,
there is an appreciable number of diffuse dwarf spheroidal galaxies (dSph).
 Measurement of radial velocities for these smooth objects of
low surface brightness is an extremely hard observational task due to
absence in them of contrast details, as well as lack of neutral hydrogen.
It is only in the closest  dSph galaxies that one can measure the
radial velocity from a globular cluster (when available) or the
brightest stars.  The majority of such ``bald'' dwarfs are
concentrated in compact groups and Virgo cluster
(Karachentseva \& Sharina, 1987). This fact suggests we can assign
to  ``bald'' dwarfs the mean distance of the group,
in whose visible perimeter they are situated. Measurements of distances
to dSphs with the HST from TRGB generaly confirm their membership
in corresponding groups.

  The total number of dSph+E companions in the complexes: Milky Way/M31,
M81, CenA/M83, and IC342/Maffei amount to 55 against 58 companions of
dIr+S types. ``Bald'' companions show stronger concentration toward
the principal galaxy as compared to gas-rich ones. Thus, the medians of
projected separations for them equal to 167 kpc (dSphs) and 210 kpc (dIrs).
One dSph object, Cetus, is at a projected distance of over 500 kpc
from the main galaxy of the group. The observed difference between the median
projected separations (167 kpc vs. 210 kps) may be considered as
insignificant. Nevertheless, among 233 galaxies having distances to
us within 5 Mpc, there are four only isolated early-type galaxies:
NGC 404 ($\Theta$ = --1.0, MD = Maffei 2), KKs3 ($\Theta$ = --0.3,
MD = NGC 1313), KK 258 ($\Theta$ = --0.9, MD = NGC 253), and Tucana
($\Theta$ = --0.1, MD = Milky Way). All of them are peculiar objects
requiring special detailed study.

  As an aside, it should be noted that the observed number of ``bald'' dwarfs
in the group tends to increase with the luminosity of bulge of the main group
member, and to decrease with crossing time of the group, which has a quite
obvious evolutionary interpretation.

  \section{ Common properties of the nearest galaxy groups}

  Some general characteristics of the discussed groups are listed in
Table 10. The following information is provided in the Table lines: (1,2)
mean group distance from us and from the LG centroid (in Mpc); when
determining the LG centroid, we assumed that the total masses of the M 31
and the Milky Way are as 5 : 4 (Karachentsev \& Makarov, 1996);
(3) distance of group center from the plane of the Local
Supercluster (in Mpc); (4,5) the known total number of members in the
group and galaxies of early-type (E+dSph); (6,7) morphological type
and absolute magnitude of the brightest member; (8) rotation velocity of
the main galaxy (in km s$^{-1}$); (9) radial velocity of the principal member
relative to the LG centroid (km s$^{-1}$); (10) mean radial velocity of
the group in the system of rest LG (km s$^{-1}$); (11) dispersion of radial
velocities in the group (in km s$^{-1}$); (12) mean projected separation of the companions
from the principal galaxy (in kpc); here the mean spatial distance of
the Milky Way companions is multiplied by ($\pi$/4) to account for projection
effects; (13) integrated blue luminosity of the group; (14,15) virial and
orbital mass estimates (in $10^{10} M_{\sun}$ units); (16,17) virial and orbital
mass-to-luminosity ratio in solar units; (18) crossing time, $\langle R_p\rangle/\sigma_v$,
in Gyr.

  As it was to be expected, two loose systems: Sculptor filament and CVnI
cloud are characterized by the greatest crossing time, which exceeds half
the Hubble time. These two systems (the last right columns in Table 10) have
obviously not reached an dynamical equilibrium, and we will no discuss them
further. Comparison of the parameters of the remaining groups allow
us to make the following statements.

  a) The centers of all the groups reside in a narrow layer
($\pm$0.33 Mpc) thick with respect to the Local supercluster plane, which
accounts for only 10\% of the volume considered.

  b) Judging by the principal characteristics: dimension, luminosity, velocity
dispersion, and content of dSphs, the Local Group is a typical representative
of nearby groups, where one main galaxy dominates.

  c) Virial/orbital mass-to-luminosity ratios for nearby groups lies within
[8 -- 88] $M_{\sun}/L_{\sun}$. Their median 29 $M_{\sun}/L_{\sun}$ is 3 -- 5
times as small as the old estimates made by Huchra \& Geller (1982) and Tully
(1987) for groups of the same luminosity.

  d) The binary structure looks to be a common feature of nearby groups.
Some paired groups: Milky Way + M 31 and M 81 + NGC 2403, manifest the
mutual
approach of their principal galaxies, and the kinematic status of the
others is open to question.

  e) The crossing times in neighboring groups are concentrated within [ 1.8 --
 5.9 ] Gyr, while the median, 2.3 Gyr, is 6 times as small as the Hubble
time, which enables these groups to be regarded as advanced in their
dynamical evolution.

\section{ Nearby groups as tools for cosmology}

  Precise measurements of distances and radial velocities for galaxies
surrounding a group permits one to determine the radius of zero- velocity
surface, $R_o$, which separates the group from the general cosmic expansion.
Determinations of $R_o$ for the six nearest groups were performed by Karachentsev
et al. (2002a,b,c, 2003a,b,c). The results are collected in Table 11.
Its first and second lines present the total mass of each group/complex
estimated by the virial theorem or from orbital motions of companions
around the principal galaxy. (As the total mass of the LG, we use the
virial/orbital mass estimates of the M31 group multiplied on the factor
1.8, which takes into account the expected mass ratio, 4:5, for the Milky
Way and the M31). The third line contains data on the turn-over
radius $R_o$ and its error. The fourth line gives the total mass calculated
from $R_o$ as $M_t = (\pi^2/8 G) R_o^3/T_o^2$, where the age of the universe
is adopted to be (13.7$\pm$0.2) Gyr (Spergel et al. 2003). The last two
lines indicate the total blue luminosity of the group/complex and the
total mass-to-luminosity ratio, respectively. Figure 1 shows the relationship
between mass estimates for the groups made by two quite independent series
of observational data: from internal motions of group members (horizontal)
and from external motions of galaxies surrounding the group. Two complexes
which are likely unbound: the Canes Venatici I cloud and the Sculptor
filament are shown by the dotted boxes.

  All four likely virialized systems: Milky Way/M31, M81/NGC2403,
CenA/M83, and IC342/Maffei manifest quite satisfactory agreement between
independent mass estimates. In two apparently unvirialized expanding
complexes (Sculptor and CVnI), their virial/orbital mass estimates turn out
to be 3 -- 8 times larger than the total masses estimated via $R_o$.
The derived total mass-to-luminosity ratios show a surprisingly
low scatter, being ranked inside [ 9 -- 37] with a median of 19
in solar units. The ratio of the sum of the total masses for the
four complexes to the sum of their total luminosity is
$21 M_{\sun}/L_{\sun}$.

  As we have already noted, all the groups under discussion are situated
in the sphere of radius 5 Mpc. From the presently available data, this
volume contains 233 galaxies. About half of them, 121/233,
are members of the LG, the M81/NGC2403 group, the CenA/M83  group,
and the IC342/Maffei group. Besides them,
62 more galaxies in the 5 Mpc- sphere (i.e. 27\%) belong to smaller
multiple systems around NGC 3109, UGC 8760, NGC 784, and UGC 3974,
and two expanding complexes in Sculptor and Canes Venatici.
Thus, the relative number of the group members in the Local volume
accounts for about 78\%, and only 22\% of nearby galaxies can be considered
to be the population of the general ``field''.

  According to the ``Catalog of Neighboring Galaxies'' (Karachentsev et al.
2004), the integrated luminosity of all galaxies within 5 Mpc is
$\Sigma L$ (5 Mpc) = $46\times 10^{10} L_{\sun}$. From the fifth line of
Table 11, the integrated luminosity of the virialized groups
is $30\times 10^{10} L_{\sun}$ or 65\% of the total luminosity of the volume.
With allowance made for the Sculptor filament and the CVnI cloud,
the fraction of local luminous matter in systems increases to 82\%.
The numbers presented reflect the known effect of segregation of galaxies
by luminosity, with the concentration of dwarf galaxies in groups less
pronounced than for giant galaxies.

  One of the remarkable properties of the spatial distribution of galaxies is
its fractality. On different scales, groups/clusters look like dense
knots in filaments which concentrate towards sheets, forming as a whole
a fractal ``cosmic web'' pattern. This general picture is valid for
the Local volume too. In the local sphere of radius 5 Mpc all the
virialized groups are situated in a layer of  $\pm$(1/3) Mpc around the
local ``pancake'', which occupies only 10\% of the sphere volume, $V$ =
523 Mpc$^3$. However, the sum of 4 group volumes with radii $R_o\sim1$ Mpc
each add to only 20 Mpc$^3$ or $\sim$4\% of the Local volume. Thus,
in the Local volume of 5 Mpc radius, only a small part ( 4\%)
is not involved in the Hubble expansion, but about 70\% of luminous
matter is concentrated in this volume. Obviously, any theory of galaxy
formation must explain these dimensionless empirical parameters
of the local web. On scales of 10 -- 100 Mpc, such an approach is being
developed by Shandarin (2004).

  Giving preference to estimates of the total mass from the pattern of the
velocity field of galaxies surrounding groups, we obtain the total mass of
six complexes (line 4 in Table 11), $\Sigma M_t = 775\times 10^{10} M_{\sun}$,
which yields the mean density of matter inside the 5 Mpc sphere
$\rho$(5 Mpc) = $1.5\times 10^{10} M_{\sun}$ Mpc$^3$. At $H_o =
72$ km s$^{-1}$ Mpc$^{-1}$, the critical density of matter, $\rho_c =
(3 H_o^2/8 \pi G$), is equal to $14.3\times 10^{10} M_{\sun}/$ Mpc$^3$.
Consequently, the mean local density of matter is only
$\rho$ ( 5 Mpc) = 0.10 $\rho_c$.

  It follows from the data of the Catalog (Karachentsev et al. 2004) that the
mean density of luminosity within 5 Mpc equals $8.7\times 10^8 L_{\sun}$/Mpc$^3$.
Comparing this with the mean luminosity density estimated from
the Sloan Digital Sky Survey (Blanton et al. 2003) and the Millenium
Galaxy Catalogue (Liske et al. 2003), we obtain a ratio
$\rho_L(5$ Mpc)/ $\rho_{L,glob} = 4.3\pm0.3$. Here we took into account that
ignoring internal extinction in galaxies in SloanDSS and MGC causes
underestimates of $\rho_{L,glob}$ by a factor of 1/3. Supposing that on
different scales the mass density is strictly proportional to
the luminosity density (no biasing), then the mean density of matter
contained in the groups leads to the value of global density
of matter of about 0.025 of the critical density.

  \section{ Peculiar motions of the groups}
Over the past few years, it has repeatedly been noted ( Sandage, 1986,
Karachentsev \& Makarov, 2001, Ekholm et al. 2001, Karachentsev et al. 2003b)
that the local Hubble flow is rather cold with a dispersion of radial
velocities less than 70 km s$^{-1}$. As can be seen from the data of
Table 10, the dispersion of virial velocities in nearby groups has
approximately the same value. For the galaxies situated outside the
groups, the estimates of $\sigma_v$ are chiefly determined by errors of
measurements of distances to galaxies, and the role played by these errors
enhances with distance. In the immediate vicinities of the LG
on scales of 2 -- 3 Mpc, the radial velocity dispersion for the
field galaxies is only 25 -- 30 km s$^{-1}$ (Karachentsev et al. 2002c).

  Apparently, it is not always easy to distinguish between true isolated
galaxies and members of loose groups. For this reason, it is interesting
to examine the behaviour of group centers, but not individual galaxies on
the Hubble diagram. Such a Hubble relation is displayed in Figure 2 using
the data of lines 2 and 10 of Table 10. The errors of measurements of
the mean velocities of the groups are not large ( $< 10$ km s$^{-1}$), and
we do not show them here. The horizontal bars correspond to the errors of
the average distance of the groups being usually $\sim$10\%.
The line corresponds to the Hubble relation with $H_o =
72$ km s$^{-1}$ Mpc$^{-1}$ and a LG zero-velocity surface at 0.95 Mpc.
Dispersion of radial velocities
obtained from these data for the centroids is 25 km s$^{-1}$
after quadratic subtracting of distance errors. The low values of ``thermal''
velocities of the galaxies in the field and the centers of the groups are
remarkably consistent with the low estimates of the mean density of matter in
the Local volume. Note that the global value $H_o = 72$ km s$^{-1}$ Mpc$^{-1}$
fits fairly the local Hubble flow.

 \section { Some other nearby groups}

Above, we considered the nearest groups only, because the quality of data
on distances of galaxies situated beyond 5 Mps is rather unsatisfactory.
For instance, the majority of galaxies in the layer $D = 5 - 10$ Mpc have no
direct individual distance estimates at all. Nevertheless, the Catalog of
Neighboring Galaxies provide us with data on tidal indices and the
names of the main disturber for each galaxy within 10 Mpc,
giving an idea of the location and population of more distant groups.
Without discussing characteristics of these groups, let us enumerate only
the most representative of them containing four or more members:
NGC 672 (5), NGC 2784 (6), NGC 3115 (7), NGC 3368/3412/3489=
Leo-I group (37), NGC 4244 (4), NGC 4594 (7), M101 (5), and NGC 6946 (7).
Here, every group is noted by the name of the brightest member, and the number
of group members with $\Theta>$ 0 is shown in brackets. Apart from the Leo-I
group, all others resemble the Local Group by a prevalence of a single
galaxy. Linear dimensions, velocity dispersions and luminosities of the
groups are similar to parameters indicated in Table 10. Only the Leo-I group
is different from the others by the presence of several giant
members of early (E,S0,Sa) types, comparable to each other in luninosity.
This group, resembling a mini-cluster, is characterized by the projected
radius of 350 kpc, the radial velocity dispersion of 130 km s$^{-1}$, and the
virial mass-to-luminosity ratio of $107 M_{\sun}/L_{\sun}$ (Karachentsev \&
Karachentseva, 2004). The galaxy groups of the Leo-I-type are much more
sparse than groups of the LG-type: in the 10 Mpc volume around us the
ratio of their numbers is approximately 1:30. Hence, the Leo-I-type
groups make a contribution of minor importance to the mean density of
matter.

  Properties of groups situated within the Local Supercluster have been
recently studied by Makarov \& Karachentsev (2000). They applyed a new
algorithm (similar to a criterion $\Theta>$ 0) to a distribution of 6320
galaxies with radial velocities $V_{LG} < 3000$ km s$^{-1}$ and selected 839 groups.
For the groups with population $N\geq 5$ the following median characteristics
were derived: the radial velocity dispersion of 86 km s$^{-1}$, the projected
harmonic radius of 250 kpc, the crossing time of 1.5 Gyr, and the virial
mass-to-luminosity ratio of $50 M_{\sun}/L_{\sun}$.
Therefore, the characteristics of the nearest groups
do not differ essentially from the main dynamical characteristics of
more representative sample covering a scale of $\sim$30 Mpc.

\section { Conclusions: the nearby groups as dark matter tracers}

  As it has been mentioned above, determinations of mass of the galaxy
clusters, made by different ways, yield mutualy concordant results with a
typical mass-to-luminosity ratio  about (250 -- 300) $M_{\sun}/L_{\sun}$ in the
$B$ band. Assuming the linear proportionality between dark and luminous
matter this leads to the global density of matter in the universe
$\Omega_m \simeq 0.27$, in agreement with WMAP (Spergel et al. 2003).
However, only (10 -- 15)\% of all galaxies are concentrated in rich clusters.
Hence, the collective input of clusters into $\Omega_m$ consist of 0.03 -- 0.04
only. Most of galaxies, about (50 -- 70)\%, are situated in groups
of different size and population, with the LG and other nearest groups
typical representatives.

Recent accurate measurements of distances to galaxies open a possibility to
measure group masses by two independent manners: via internal and via
external galaxy motions. Applied to the well studied neighboring
groups, both the methods manifest good mutual agreement. However the
mass-to-luminosity ratios for groups turn out to be one order less
than for clusters. It means that the hypothesis assuming linear
proportionality of dark and luminous matter on different scales 
is not valid.

  Since the ``external'' method estimates the total mass of group on a
scale of $R_o \sim 1$ Mpc, but the ``internal'' method yields virial/orbital
mass estimate on a lower scale, $\langle R_p\rangle \simeq 200$ kpc, then the
dark matter in groups is tightly tied to the luminous one. The
case of the nearest group around Andromeda shows that the total
mass of the M31 halo, $75\times 10^{10} M_{\sun}$ is reached already on the scales
of (50 -- 125) kpc, i.e. one order less than the group turn-over radius $R_o$.

  As it is seen from Figure 2, centroids of the nearby groups have a low
velocity dispersion with respect to the Hubble flow that is an independent
evidence of the low mean density of matter in the Local volume
(Governato et al. 1997). The principal galaxies in nearby groups (shown
by crosses in Figure 2) have small peculiar velocities too. Only in the
M 81 group do we find a significant peculiar velocity of the main member.
(The situation in the Maffei group is complicated by 
heavy Galactic extinction, as well as confusion of HI emission from
the group members with the Galactic HI emission). An apparent reason of
the high peculiar velocity of the M 81 is the presence nearby M 81 of another
bright galaxy, M 82, which has a peculiar velocity of opposite sign (see
Table 3). Moreover, the peculiar velocities of M 81 and M 82 are inversly
proportional to their luminosities. Consequently, the galaxies M 81 and
M 82 are moving in the group as heavy bodies driven by the law
of conservation of motion, rather than as test particles inside the common
smooth potential well.

  Supposing the dark matter in groups follows tightly the luminous matter, we
can estimate the total contribution of groups to the mean density of matter.
Based on the average weighted values: $\langle M_{vir}/L_B\rangle = 31$,
$\langle M_{orb}/L_B\rangle = 34$, and $\langle M_t/L_B\rangle = 21$ in solar units, we derive the total
contribution of groups of galaxies to the mean density of matter to
be (0.03 -- 0.04) in critical density units.

  Recently, Guzik \& Seljak (2002) and Hoekstra et al. (2004) determined
parameters of dark halos for field galaxies and members of loose groups
from weak lensing in the Sloan DSS and in the Red-Sequence Cluster Survey.
For a galaxy with the mean luminosity of $L_B = 2\times 10^{10} L_{\sun}  (H_o =
72$ km s$^{-1}$ Mpc$^{-1}$) they derived a characteristic
ratio $M/L_B$ = 41 $M_{\sun}/L_{\sun}$
on a scale of $\sim$250 kpc (after correction for internal extinction).
This independent result agrees well with our data.

  Thus, we may conclude that groups and clusters of galaxies
produce approximately the same contribution to the mean density of
matter. Their combined contribution is about 0.06 -- 0.08 in the units
of critical density or about (1/5 -- 1/3) with respect to $\Omega_m = 0.27$.
To avoid a contradiction with the global matter density $\Omega_m = 0.27$,
derived from the WMAP, we need to assume the existence of another
(non-baryonic) component of dark matter with the mean
density $\Omega_{dm2} \simeq 0.20$. The DM2-component may be
distributed in space either homogeneously likes a dark cosmic ``ocean'',
or consists of many low mass halos without gas and stars,
as assumed by Tully et al. (2002) and Tully (2004). Obviously,
the presence of the DM2- component does not significantly affect the
kinematics of even loose groups, because it contributes a small fraction
(about 10\% within $R_o = 1$ Mpc and about 0.1\% within $R_{vir}= 0.2$ Mpc)
into the integrated mass of groups.

\acknowledgements{
I thank B. Tully and P.J.E. Peebles for useful discussions.

Support for this work was provided by NASA through grant GO--08601.01--A from
the Space Telescope Science Institute, which is operated by the Association
of Universities for Research in Astronomy, Inc., under NASA contract
NAS5--26555.
 This work was also supported by RFFI grant 04--02--16115 and
DFG-RFBR grant 02--02--04012. This search has made use of the NASA/IPAC
Extragalactic Database (NED).}

{}
\newpage
\figcaption [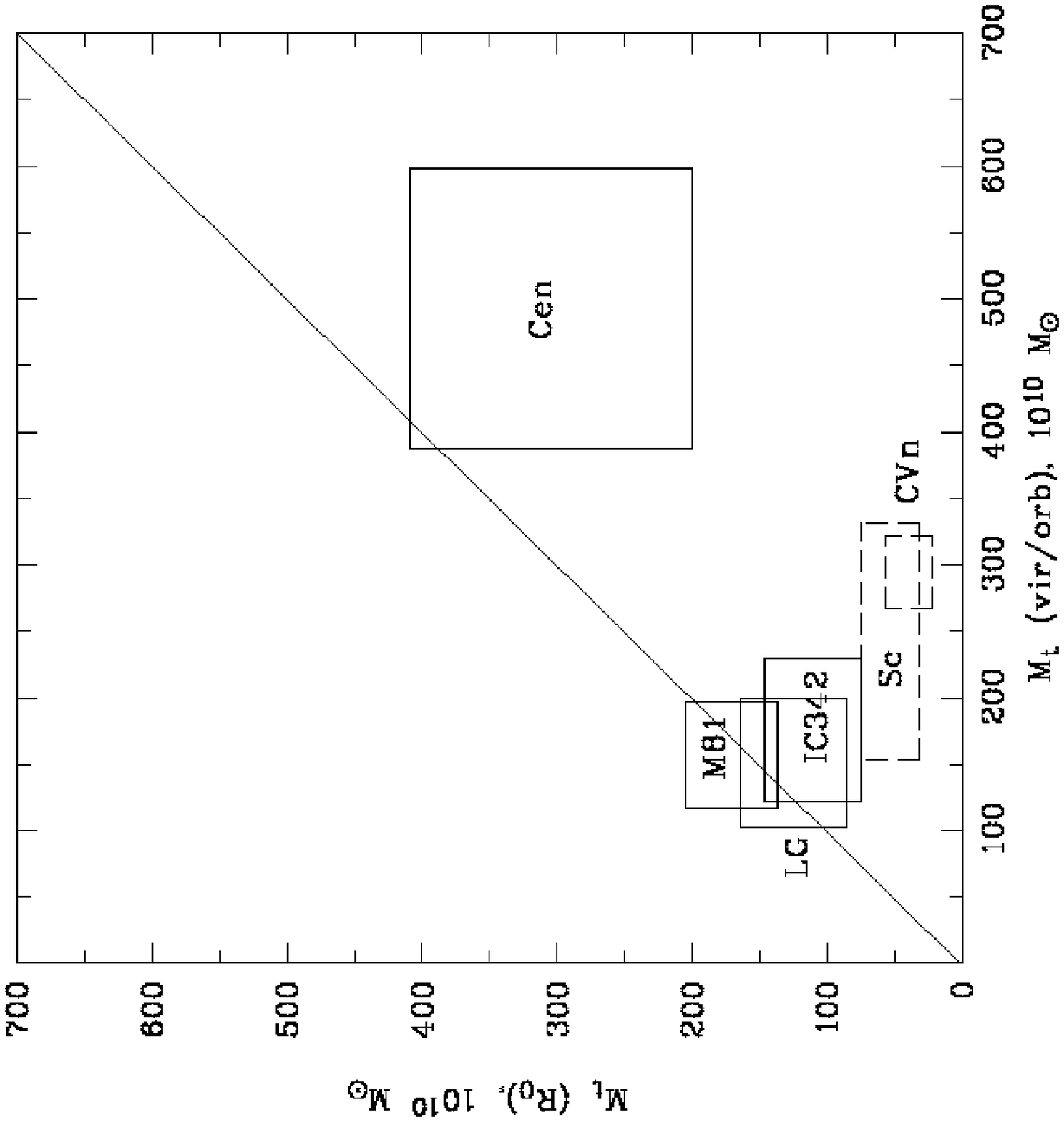]{The total mass estimates derived from the zero-velocity
surface at R$_o$ (vertical) versus the virial/orbital masses for the nearest
galaxy complexes: Milky Way/M31 (the Local Group), M81/NGC2403,
IC342/Maffei, CenA/M83. Two regions that are likely unbound:
the Canes Venatici I cloud and the Sculptor filament are shown by
the dotted boxes.}

\figcaption [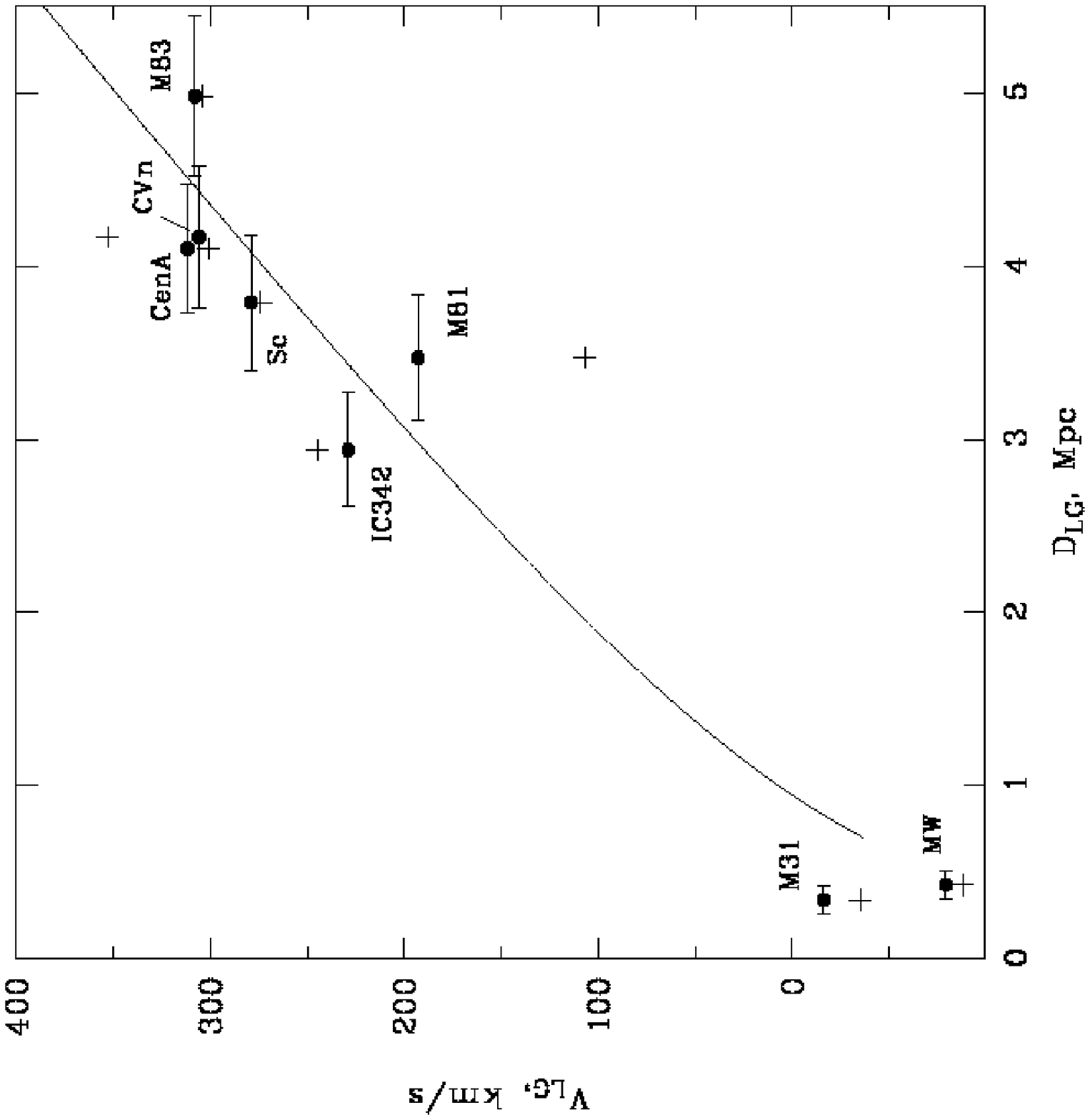]{ The Hubble diagram for the centroids of the nearest groups.
The distances and radial velocities are shown with respect to the Local Group
centroid. The line corresponds to the Hubble relation with $H_o =
72$ km s$^{-1}$ Mpc$^{-1}$ and a LG zero-velocity surface at 0.95 Mpc.
The brightest members of the groups are indicated by crosses.}
\begin{deluxetable}{lrrrrlr}
\tablewidth{0pc}
\tablecaption{The Milky Way group.\label{table1}}
\tablehead{
  \colhead{Name}&
  \colhead{Ty} &
  \colhead{$V_{MW}$}  &
  \colhead{$V_{LG}$}&
  \colhead{$M_B$}&
  \colhead{$D$}&
  \colhead{$\Theta$}\\
  \colhead{}&
  \colhead{} &
  \colhead{km s$^{-1}$}  &
  \colhead{km s$^{-1}$}&
  \colhead{mag}&
  \colhead{Mpc}&
  \colhead{}\\
}
\startdata
 Milky Way  &   4  &  0 &  $-$88  &  $-$20.80 & 0.01~~cep & 2.5  \\
 Sgr dSph   &  $-$3  &171 &  161  &  $-$12.67 & 0.024 rgb & 5.6  \\
 LMC        &   9  & 84 &   28  &  $-$17.93 & 0.050 cep & 3.6  \\
 SMC        &   9  & 17 &  $-$22  &  $-$16.35 & 0.063 cep & 3.5  \\
 Ursa Min    &  $-$3  &$-$85 &  $-$44  &  $-$ 7.13 & 0.063 rgb & 3.3   \\
 Draco      &  $-$3  &$-$98 &  $-$48  &  $-$ 8.74 & 0.079 rgb & 3.0 \\
 Sex dSph   &  $-$3  & 74 &    8  &  $-$ 7.98 & 0.086 rgb & 2.8 \\
 Sculptor   &  $-$3  & 77 &   96  &  $-$ 9.77 & 0.088 rgb & 2.8  \\
 Carina     &  $-$3  &  7 &  $-$53  &  $-$ 8.97 & 0.10~~rgb & 2.7 \\
 Fornax     &  $-$3  &$-$36 &  $-$32  &  $-$11.50 & 0.14~~rgb & 2.3 \\
 Leo II     &  $-$3  & 22 &  $-$18  &  $-$ 9.23 & 0.21~~rgb & 1.7 \\
 Leo I      &  $-$3  &177 &  128  &  $-$10.97 & 0.25~~rgb & 1.5 \\
 Phoenix    &  $-$1& $-$103 & $-$106  &  $-$10.22 & 0.44~~rgb & 0.8 \\
 NGC 6822   &  10  & 44 &   64  &  $-$15.22 & 0.50~~cep & 0.6 \\
 Leo A      &  10  &$-$15 &  $-$40  &  $-$11.36 & 0.69~~rgb & 0.2 \\
\hline
 Tucana     &  $-$2  & 35 &   9   &  $-$ 9.16 & 0.88~~rgb &$-$0.1  \\
 SagDIG     &  10  &  9 &  23   &  $-$11.49 & 1.04~~rgb &$-$0.3   \\
 Sex A      &  10  &163 &  94   &  $-$13.95 & 1.32~~cep &$-$0.6     \\
 Sex B      &  10  &168 & 111   &  $-$13.96 & 1.36~~rgb &$-$0.7       \\
\hline
\enddata
\end{deluxetable}

\begin{deluxetable}{lrrrrcr}
\tablewidth{0pc}
\tablecaption{The Andromeda group\label{table1_1}}
\tablehead{
  \colhead{Name}&
  \colhead{Type} &
  \colhead{$V_{LG}$}&
  \colhead{$M_B$}&
  \colhead{$D$}&
  \colhead{$\theta$}&
  \colhead{$\Theta$}\\
  \colhead{}&
  \colhead{} &
  \colhead{km s$^{-1}$}  &
  \colhead{mag}&
  \colhead{Mpc}&
  \colhead{deg}&
  \colhead{} \\
}
\startdata
 M~31        &   3  & $-$35  & $-$21.58  & 0.77 cep &   $\phantom{0}$0.0  & 4.6 \\
 NGC~221=M~32   &  $-$5  & 121  & $-$15.96  & 0.77 rgb &   $\phantom{0}$0.40 & 6.8 \\
 NGC~205       &  $-$5  &  24  & $-$16.15  & 0.83 rgb &   $\phantom{0}$0.68 & 3.7 \\
 And IX     &  $-$3  &   $-$  & $-$ 7.5   & 0.79 rgb & $\phantom{0}$2.60 & 3.8 \\
 And I      &  $-$3  &$-$120  & $-$10.87  & 0.81 rgb & $\phantom{0}$3.31 & 3.7 \\
 And III    &  $-$3  & $-$92  & $-$ 9.30  & 0.76 rgb & $\phantom{0}$4.98 & 3.5 \\
 NGC~185       &  $-$3  &  73  & $-$14.76  & 0.62 rgb &   $\phantom{0}$7.10 & 2.3 \\
 NGC~147       &  $-$3  &  85  & $-$14.79  & 0.76 rgb &   $\phantom{0}$7.43 & 3.0 \\
 And V      &  $-$3  &$-$143  & $-$ 8.41  & 0.81 rgb & $\phantom{0}$8.03 & 2.8 \\
 And II     &  $-$3  &  46  & $-$ 9.33  & 0.68 rgb &  10.31 & 2.4 \\
 M~33        &   5  &  36  & $-$18.87  & 0.85 cep &  14.72 & 2.0 \\
 Cas dSph   &  $-$3  & $-$ 5  & $-$11.67  & 0.79 rgb &  16.17 & 2.0 \\
 IC 10      &  10  & $-$60  & $-$15.57  & 0.66 cep &  18.42 & 1.8 \\
 Peg dSph   &  $-$3  & $-$94  & $-$10.80  & 0.82 cep &  19.77 & 1.7 \\
 LGS 3      &  $-$1  & $-$74  & $-$ 7.96  & 0.62 rgb &  19.87 & 1.7 \\
 Pegasus    &  10  &  60  & $-$11.47  & 0.76 rgb &  31.02 & 1.2 \\
 IC 1613    &  10  & $-$89  & $-$14.51  & 0.73 cep &  39.37 & 0.9 \\
 Cetus      &  $-$2  &   $-$  & $-$10.18  & 0.78 rgb &  52.4  & 0.5 \\
 WLM        &   9  & $-$10  & $-$13.59  & 0.92 rgb &  57.5  & 0.3 \\
\hline
 DDO 210    &  10  &  13  & $-$11.09  & 0.94 rgb &  76.6  &$-$0.1 \\
 KKH 98     &  10  & 151  & $-$10.78  & 2.45 rgb &  11.23 &$-$0.7 \\
 KKR 25     &  10  &  68  & $-$ 9.94  & 1.86 rgb &  74.3  &$-$0.7 \\
 KK 230     &  10  & 126  & $-$ 8.55  & 1.90 rgb & 101.2  &$-$1.0 \\
\hline
\enddata
\end{deluxetable}
\begin{deluxetable}{lrrrrlrl}
\tablewidth{0pc}
\tablecaption{The  M81 group.\label{table3}}
\tablehead{
  \colhead{Name}&
  \colhead{Ty} &
  \colhead{$V_{LG}$}  &
  \colhead{$R_p$}&
  \colhead{$M_B$}&
  \colhead{$D$}&
  \colhead{$\Theta$}&
  \colhead{Note}\\
  \colhead{}&
  \colhead{} &
  \colhead{km s$^{-1}$}  &
  \colhead{kpc}&
  \colhead{mag}&
  \colhead{Mpc}&
  \colhead{}&
  \colhead{}\\
}
\startdata
 M~81    &   3   &  107   &  0  & $-$21.06 & 3.63 cep & 2.2 &              \\
 Ho IX  &  10   &  188   & 11  & $-$13.68 & 3.7  mem & 3.3 &              \\
 BK3N   &  10   &  101   & 11  & $-$ 9.59 & 4.02 rgb & 1.0 &              \\
 A0952  &  10   &  243   & 18  & $-$11.51 & 3.87 rgb & 1.9 &MD= N3077     \\
 KDG~61  &  $-$1   &   23   & 33  & $-$12.85 & 3.60 rgb & 3.9 &              \\
 M~82    &   8   &  347   & 39  & $-$19.63 & 3.53 rgb & 2.7 &              \\
 NGC~3077  &  10   &  153   & 49  & $-$17.76 & 3.82 rgb & 1.9 &              \\
 FM1    &  $-$3   &   $-$    & 62  & $-$10.48 & 3.42 rgb & 1.8 &MD= M82       \\
 BK5N   &  $-$3   &   $-$    & 73  & $-$10.61 & 3.78 rgb & 2.4 &MD= N3077     \\
 IKN    &  $-$3   &   $-$    & 84  & $-$11.44 & 3.7  mem & 2.7 &  \\
 NGC~2976  &   5   &  139   & 87  & $-$17.10 & 3.56 rgb & 2.7 &              \\
 KDG~64  &  $-$3   &   $-$    &103  & $-$12.57 & 3.70 rgb & 2.5 &              \\
 KK~77   &  $-$3   &   $-$    &104  & $-$12.03 & 3.48 rgb & 2.0 &              \\
 F8D1   &  $-$3   &   $-$    &121  & $-$12.59 & 3.77 rgb & 2.0 &              \\
 HIJASS &  13   &  187   &147  & $-$ 7.9: & 3.7  mem & 2.2 &              \\
 Ho I   &  10   &  291   &156  & $-$14.49 & 3.84 rgb & 1.5 &              \\
 KDG~63  &  $-$3   &    0   &169  & $-$12.12 & 3.50 rgb & 1.8 &              \\
 HS~117  &  10   &  116   &190  & $-$11.83 & 3.7  mem & 1.9 &   \\
 IC~2574 &   9   &  197   &193  & $-$17.46 & 4.02 rgb & 0.9 &              \\
 DDO~78  &  $-$3   &  191   &201  & $-$12.17 & 3.72 rgb & 1.8 &              \\
 DDO~82  &   9   &  207   &214  & $-$14.63 & 4.00 rgb & 0.9 &              \\
 BK6N   &  $-$3   &   $-$    &304  & $-$11.08 & 3.85 rgb & 1.1 &              \\
 KDG~73  &  10   &  263   &321  & $-$10.83 & 3.70 rgb & 1.3 &              \\
 KKH~57  &  $-$3   &   $-$    &373  & $-$10.19 & 3.93 rgb & 0.7 &              \\
 DDO~53  &  10   &  151   &519  & $-$13.37 & 3.56 rgb & 0.7 &              \\
 KDG~52  &  10   &  268   &506  & $-$11.49 & 3.55 rgb & 0.7 &              \\
 Ho II  &  10   &  311   &530  & $-$16.72 & 3.39 rgb & 0.6 &              \\
 UGC~4483  &  10   &  304   &436  & $-$12.73 & 3.21 rgb & 0.5 &              \\
 NGC~2403  &   6   &  268   &854  & $-$19.29 & 3.30 cep & 0.0 &              \\
\hline
 UGC~6456  &  10   &   89   &799  & $-$14.03 & 4.34 rgb &$-$0.3 &              \\
 NGC~4236  &   8   &  160   &776  & $-$18.59 & 4.45 rgb &$-$0.4 &              \\
 DDO~44  &  $-$3   &   $-$    &833  & $-$12.07 & 3.19 rgb & 1.7 &MD=N2403      \\
 NGC~2366  &  10   &  253   &809  & $-$16.02 & 3.19 rgb & 1.0 &MD=N2403     \\
 UGC~7242  &  10   &  213   &840  & $-$13.65 & 4.3  mem & 0.4 &MD=N4236      \\
 DDO~165 &  10   &  196 &1075  & $-$15.09 & 4.57 rgb & 0.0 &MD=N4236      \\
        \hline
\enddata
\end{deluxetable}

\begin{deluxetable}{lrrrrlrl}
\tablewidth{0pc}
\tablecaption{The Centaurus A group\label{table4}}
\tablehead{
  \colhead{Name}&
  \colhead{Ty} &
  \colhead{$V_{LG}$}  &
  \colhead{$R_p$}&
  \colhead{$M_B$}&
  \colhead{$D$}&
  \colhead{$\Theta$}&
  \colhead{Note}\\
  \colhead{}&
  \colhead{} &
  \colhead{km s$^{-1}$}  &
  \colhead{kpc}&
  \colhead{mag}&
  \colhead{Mpc}&
  \colhead{}&
  \colhead{}\\
}
\startdata
 NGC~5128       &  $-$2   &301   &  0   &$-$20.77  &3.66 rgb  &0.6 &MD=N4945   \\
 KKs~55       &   $-$3  &  $-$   &  42  & $-$ 9.91 & 3.6  mem & 3.1&           \\
 KK~197       &   $-$3  &  $-$   &  50  & $-$12.76 & 3.6  mem & 3.0&            \\
 ESO~324-024    &   10  & 270  & 102  & $-$15.45 & 3.73 rgb & 2.4&             \\
 KK~196       &   10  & 490  & 138  & $-$12.00 & 3.6  mem & 2.2&             \\
 NGC~5237       &   $-$3  & 131  & 143  & $-$15.00 & 3.6  mem & 2.1&     \\
 KK~203       &   $-$3  &  $-$   & 150  & $-$10.22 & 3.6  mem & 2.1&              \\
 KK~189       &   $-$3  &  $-$   & 167  & $-$10.52 & 3.6  mem & 2.0&             \\
 ESO~269-66,KK~190 & $-$5  & 528  & 184  & $-$13.56 & 3.54 sbf & 1.7&              \\
 KKs~57       &   $-$3  &  $-$   & 190  & $-$10.07 & 3.6  mem & 1.8&               \\
 KK~213       &   $-$3  &  $-$   & 214  & $-$ 9.72 & 3.63 rgb & 1.7&       \\
 KK~211       &   $-$5  &  $-$   & 235  & $-$11.93 & 3.58 rgb & 1.5&        \\
 ESO~325-011    &   10  & 307  & 242  & $-$14.05 & 3.40 rgb & 1.1&         \\
 KK~217       &   $-$3  &  $-$   & 291  & $-$10.87 & 3.84 rgb & 1.1&          \\
 ESO~269-058    &   10  & 142  & 305  & $-$14.95 & 3.6  mem & 1.9& MD=N4945  \\
 KKs~53,Cen7  &   $-$3  &  $-$   & 312  & $-$10.86 & 3.6  mem & 1.2&            \\
 ESO~269-37,KK~179 & $-$3  &  $-$   & 334  & $-$12.02 & 3.48 rgb & 1.6& MD=N4945  \\
 NGC~5206       &   $-$3  & 322  & 342  & $-$16.66 & 3.6  mem & 1.1& MD=N4945   \\
 Cen6,KK~182  &   10  & 360  & 350  & $-$11.89 & 3.6  mem & 1.2&      \\
 KK~221       &   $-$3  &  $-$   & 364  & $-$10.60 & 3.98 rgb & 0.6&            \\
 CenN        &   $-$3  &  $-$   & 384  & $-$10.89 & 3.6  mem & 0.9&             \\
 HIPASS~1351  &   10  & 292  & 388  & $-$10.90 & 3.6  mem & 0.9&            \\
 NGC~5102       &    1  & 230  & 410  & $-$18.08 & 3.40 rgb & 0.7&            \\
 HIPASS~1348  &   10  & 347  & 426  & $-$11.21 & 3.6  mem & 0.8&            \\
 NGC~4945       &    6  & 296  & 468  & $-$20.51 & 3.6  mem & 0.7&            \\
 KKs~51       &   $-$3  &  $-$   & 476  & $-$11.46 & 3.6  mem & 0.7&            \\
 KKs~58       &   $-$3  &  $-$   & 492  & $-$10.64 & 3.6  mem & 0.6&       \\
 ESO~384-016    &   10  & 350  & 624  & $-$13.06 & 3.72 sbf & 0.3&        \\
\hline
 ESO~219-010    &  $-$3   & $-$    &557   &$-$12.70  &4.28 sbf  &0.1 &MD=N4945\\
 PGC~51659      &  10   &171   &736   &$-$11.83  &3.58 rgb  &0.1 &        \\
 ESO~321-014    &  10   &337   &914   &$-$12.70  &3.19 rgb &$-$0.3 &        \\
        \hline
\enddata
\end{deluxetable}
\begin{deluxetable}{lrrrrlrl}
\tablewidth{0pc}
\tablecaption{The M83 group\label{table4_1}}
\tablehead{
  \colhead{Name}&
  \colhead{Ty} &
  \colhead{$V_{LG}$}  &
  \colhead{$R_p$}&
  \colhead{$M_B$}&
  \colhead{$D$}&
  \colhead{$\Theta$}&
  \colhead{Note}\\
  \colhead{}&
  \colhead{} &
  \colhead{km s$^{-1}$}  &
  \colhead{kpc}&
  \colhead{mag}&
  \colhead{Mpc}&
  \colhead{}&
  \colhead{}\\
}
\startdata

 NGC~5236,M~83 &   5  & 304  &   0   & $-$20.43 & 4.47 cep & 0.8& MD=N5264   \\
 KK~208     &  $-$3  &  $-$   &  25   & $-$14.24 & 4.68 rgb & 1.6&            \\
 PGC~47885    &  10  & 360  &  40   & $-$12.98 & 5.0  h   & 0.4&            \\
 ESO~444-078  &  10  & 363  &  51   & $-$13.01 & 4.6  mem & 2.1&            \\
 NGC~5264     &  10  & 268  &  80   & $-$15.90 & 4.53 rgb & 2.6&            \\
 IC~4316    &  10  & 382  &  95   & $-$13.90 & 4.41 rgb & 2.4&            \\
 ESO~444-084  &  10  & 380  & 145   & $-$13.56 & 4.61 rgb & 1.7&            \\
 NGC~5253     &   8  & 190  & 154   & $-$17.38 & 4.00 cep & 0.5&            \\
 KK~218     &  $-$3  &  $-$   & 167   & $-$10.97 & 4.6  mem & 1.6&            \\
 IC~4247    &  10  & 195  & 181   & $-$14.18 & 4.6  mem & 1.5&            \\
 KK~200     &   9  & 264  & 230   & $-$11.96 & 4.63 rgb & 1.2&            \\
 DEEP~1337-33& 10  & 371  & 279   & $-$11.18 & 4.51 rgb & 1.2&            \\
 KKs~54     &  $-$3  &  $-$   & 308   & $-$10.47 & 4.6  mem & 1.0&            \\
 KK~198     &  $-$3  &  $-$   & 379   & $-$10.96 & 4.6  mem & 0.8&            \\
\hline
 KK~195     &  10  & 338  & 302   & $-$11.76 & 5.22 rgb &$-$0.2&            \\
 ESO~381$-$020  &  10  & 332  & 911   & $-$14.15 & 4.6  h   &$-$0.3&            \\
 HIPASS~1337& 10  & 258  & 792   & $-$12.27 & 4.90 rgb &$-$0.3&            \\
 NGC~5408     &  10  & 288  &1002   & $-$16.50 & 4.81 rgb &$-$0.5&            \\
 ESO~381-018  &  10  & 353  & 990   & $-$13.00 & 4.9  h   &$-$0.6&       \\

        \hline
\enddata
\end{deluxetable}

\begin{deluxetable}{lrrrllr}
\tablewidth{0pc}
\tablecaption{The IC 342  group\label{table6}}
\tablehead{
  \colhead{Name}&
  \colhead{Ty} &
  \colhead{$V_{LG}$}  &
  \colhead{$R_p$}&
  \colhead{$M_B$}&
  \colhead{$D$}&
  \colhead{$\Theta$} \\
  \colhead{}&
  \colhead{} &
  \colhead{km s$^{-1}$}  &
  \colhead{kpc}&
  \colhead{mag}&
  \colhead{Mpc}&
  \colhead{}\\

}
\startdata

 IC 342  &     5  &  245  &    0  & $-$20.69 & 3.28 cep & $-$0.1  \\
 KK~35    &    10  &  149  &   15  & $-$14.30 & 3.16 rgb &  2.4  \\
 UA~86    &     8  &  275  &   91  & $-$18.06 & 3.12 rgb &  1.9  \\
 NGC~1560   &     8  &  171  &  311  & $-$16.87 & 3.45 rgb &  1.0  \\
 CamB    &    10  &  266  &  370  & $-$11.85 & 3.34 rgb &  1.0  \\
 CamA    &    10  &  164  &  325  & $-$14.06 & 3.93 rgb &  0.1  \\
 Cas1    &    10  &  283  &  529  & $-$16.70 & 3.3  mem &  0.5  \\
 UGCA~105   &     9  &  279  &  615  & $-$16.81 & 3.15 rgb &  0.3  \\
\hline
 KKH~37   &    10  &  204  &  932  & $-$11.55 & 3.34 rgb & $-$0.3  \\
        \hline
\enddata
\end{deluxetable}

\begin{deluxetable}{lrrrllr}
\tablewidth{0pc}
\tablecaption{The Maffei group\label{table7}}
\tablehead{
  \colhead{Name}&
  \colhead{Ty} &
  \colhead{$V_{LG}$}  &
  \colhead{$R_p$}&
  \colhead{$M_B$}&
  \colhead{$D$}&
  \colhead{$\Theta$} \\
  \colhead{}&
  \colhead{} &
  \colhead{km s$^{-1}$}  &
  \colhead{kpc}&
  \colhead{mag}&
  \colhead{Mpc}&
  \colhead{} \\
}
\startdata

 Maffei~2   &   4  &  212  &    0 &  $-$20.15&   2.8  tf &  1.4 \\
 Maffei~1   &  $-$3  &  246  &   36 &  $-$18.97&   3.01 fj &  1.7 \\
 MB~1       &   7  &  421  &   49 &  $-$14.81&   3.0  mem&  1.7 \\
 MB~3       &  10  &  280  &   74 &  $-$13.65&   3.0  mem&  1.6 \\
 Dwing~2    &  10  &  316  &   88 &  $-$14.55&   3.0  mem&  1.6 \\
 Dwing~1    &   3  &  333  &  107 &  $-$18.78&   2.8  tf &  2.5 \\
 KKH~12     &  10  &  303  &  146 &  $-$13.03&   3.0  mem&  1.4 \\
 KKH~11     &  10  &  308  &  225 &  $-$13.35&   3.0  mem&  1.0 \\
\hline
 KKH~6      &  10  &  270  &  630 &  $-$12.42&   3.8  rgb& $-$0.8   \\
       \hline
\enddata
\end{deluxetable}

\begin{deluxetable}{lrrrllr}
\tablewidth{0pc}
\tablecaption{The Sculptor filament\label{table8}}
\tablehead{
  \colhead{Name}&
  \colhead{Ty} &
  \colhead{$V_{LG}$}  &
  \colhead{$R_p$}&
  \colhead{$M_B$}&
  \colhead{$D$}&
  \colhead{$\Theta$} \\
  \colhead{}&
  \colhead{} &
  \colhead{km s$^{-1}$}  &
  \colhead{kpc}&
  \colhead{mag}&
  \colhead{Mpc}&
  \colhead{} \\
}
\startdata
 NGC~253    &  5  &   274  &    0   & $-$21.37 & 3.94 rgb & 0.3 \\
 DDO 6    & 10  &   348  &  287   & $-$12.50 & 3.34 rgb & 0.5 \\
 NGC 247    &  7  &   215  &  302   & $-$18.81 & 4.09 tf  & 1.3 \\
 Sc 22    & $-$3  &    $-$   &  361 & $-$10.45 & 4.21 rgb & 0.9 \\
 ESO~540-032 & $-$3  &    $-$   &  362 & $-$11.32 & 3.42 rgb & 0.6 \\
 KDG~2     & $-$1  &    $-$   &  482 & $-$11.39 & 3.40 rgb & 0.4 \\
\hline
 NGC~7793    &  7  &   252  &  892   & $-$18.53 & 3.91 rgb & 0.1 \\
 DDO 226  & 10  &   408  &  281   & $-$14.17 & 4.92 rgb &$-$0.1 \\
 NGC 625    &  9  &   335  & 1292   & $-$16.53 & 4.07 rgb &$-$0.4 \\
 NGC 59     & $-$3  &   431  &  570   & $-$15.74 & 5.30 sbf &$-$0.6 \\
 ESO~245-05  &  9  &   308  & 1480   & $-$15.59 & 4.43 rgb &$-$0.7 \\
       \hline
\enddata
\end{deluxetable}

\begin{deluxetable}{lrrrllr}
\tablewidth{0pc}
\tablecaption{The Canes Venatici I cloud \label{table9}}
\tablehead{
  \colhead{Name}&
  \colhead{Ty} &
  \colhead{$V_{LG}$}  &
  \colhead{$R_p$}&
  \colhead{$M_B$}&
  \colhead{$D$}&
  \colhead{$\Theta$} \\
  \colhead{}&
  \colhead{} &
  \colhead{km s$^{-1}$}  &
  \colhead{kpc}&
  \colhead{mag}&
  \colhead{Mpc}&
  \colhead{} \\
}
\startdata

 NFC 4736,M~94 &  2  &   353  &    0  &$-$19.83  &4.66 rgb  &$-$0.5 \\
 KK~160     & 10  &   346  &  205  &$-$11.52  &4.8  h    & 1.0 \\
 IC~3687     & 10  &   385  &  240  &$-$14.64  &4.57 rgb  & 1.1 \\
 IC~4182     &  9  &   356  &  340  &$-$16.40  &4.70 cep  & 0.6 \\
 NGC~4449     &  9  &   249  &  367  &$-$18.27  &4.21 rgb  & 0.0 \\
 DDO~126    & 10  &   231  &  435  &$-$14.38  &4.87 rgb  & 0.1 \\
 KK~166     & $-$3  &   $ -$   &  440  &$-$10.82  &4.74 rgb  & 0.3 \\
 DDO~168    & 10  &   273  &  490  &$-$15.28  &4.33 rgb  & 0.0 \\
 NGC~4244     &  6  &   255  &  560  &$-$18.60  &4.49 rgb  & 0.0 \\
\hline
 NGC~5229     &  7  &   460  &  735  &$-$14.60  &5.1  bs   &$-$0.6  \\
\hline
\enddata
\end{deluxetable}

\begin{deluxetable}{lccccccc|cc}
\tablewidth{0pc}
\tablecaption{Basic properties of the nearest galaxy groups.\label{table}}
\tablehead{
  \colhead{Parameter}&
  \colhead{M.Way} &
  \colhead{M31}  &
  \colhead{M81}&
  \colhead{CenA}&
  \colhead{M83}&
  \colhead{IC342}&
  \colhead{Maffei}&
  \colhead{Sc}&
  \colhead{CVnI}\\
}
\startdata
$D_{MW}$, Mpc    &  0.01 &  0.77  & 3.63 &  3.66  & 4.56  & 3.28  & 3.01: & 3.94 &  4.09 \\
             &       &        &      &        &       &       &       &      &       \\
$D_{LG}$, Mpc    &  0.43 &  0.34  & 3.47 &  4.10  & 4.98  & 2.94  & 2.67: & 3.79 &  4.17 \\
           &       &        &      &        &       &       &       &      &       \\
$SGZ$, Mpc     &  0.00 &  0.07  & 0.04 & $-$0.33  & 0.08  & 0.02  & 0.08  &$-$0.34 &  0.77 \\
             &       &        &      &        &       &       &       &      &       \\
$N_{tot}$        &   15  &   19   &  29  &   28   &  14   &  8    &  8:   &  6   &   9   \\
             &       &        &      &        &       &       &       &      &       \\
$N_{E+dSph}$   &   10  &   13   &  11  &   18   &   4   &  0    &  1:   &  3   &   1   \\
             &       &        &      &        &       &       &       &      &       \\
Ty(1)      &    4  &    3   &   3  &   $-$2   &   5   &  5    &  4    &  5   &   2   \\
             &       &        &      &        &       &       &       &      &       \\
$M_B$(1), mag  & $-$20.80& $-$21.58 &$-$21.06& $-$20.77 &$-$20.43 &$-$20.69 &$-$20.15 &$-$21.37& $-$19.83\\
             &       &        &      &        &       &       &       &      &       \\
$V_m$(1), km s$^{-1}$ &   220 &   255  &  232 &   398  &  211  &  162  &  163  &  199 &   164 \\
             &       &        &      &        &       &       &       &      &       \\
$V_{LG}$(1), km s$^{-1}$&   $-$88 &   $-$35  &  107 &   301  &  304  &  245  &  212  &  274 &   353 \\
             &       &        &      &        &       &       &       &      &       \\
$\langle V_{LG}\rangle$, km s$^{-1}$ &   $-$79 &   $-$16  &  193 &   312  &  308  &  229  &  302  &  279 &   306 \\
             &       &        &      &        &       &       &       &      &       \\
$\sigma_v$, km s$^{-1}$ &    76 &    77  &   91 &   105  &   71  &   54  &   59  &   54 &    56 \\
             &       &        &      &        &       &       &       &      &       \\
$\langle R_p\rangle$, kpc   &   155&   254  &  211 &   290  &  164  &  322  &  104  &  359 &   385  \\
             &       &        &      &        &       &       &       &      &        \\
$L_B, 10^{10}L_{\sun}$ &  3.28 &  6.83  & 6.11 &  5.55  & 2.31  & 3.21  & 2.69  & 5.58 &  2.00  \\
             &       &        &      &        &       &       &       &      &         \\
$M_{vir}, 10^{10}M_{\sun}$ &   93  &   57   & 117  &  489   & 109   &  57   &  65   & 332  &   267  \\
             &       &        &      &        &       &       &       &      &        \\
$M_{orb}, 10^{10}M_{\sun}$ &   96  &  111   & 197  &  288   & 100   &  95   & 135   & 153  &   322  \\
             &       &        &      &        &       &       &       &      &        \\
$M_{vir}/L$      &   28  &    8   &  19  &   88   &  47   &  18   &  24   &  60  &   133  \\
             &       &        &      &        &       &       &       &      &        \\
$M_{orb}/L$      &   29  &   16   &  32  &   52   &  43   &  30   &  50   &  28  &   161  \\
             &       &        &      &        &       &       &       &      &        \\
$T_{cross}$, Gyr &   2.1 &   3.3  &  2.3 &   2.8  &  2.3  & 5.9   & 1.8   & 6.6  &   6.9   \\
\hline
\enddata
\end{deluxetable}

\begin{deluxetable}{lcccccc}
\tablewidth{0pc}
\tablecaption{Total mass estimates for the neighboring galaxy groups.\label{Table11}}
\tablehead{
  \colhead{Parameter}&
  \colhead{M.Way/M31} &
  \colhead{M81/N2403}&
  \colhead{CenA/M83}&
  \colhead{IC342/Maff}&
  \colhead{Sculptor}&
  \colhead{CVn I}\\
}
\startdata
 $\Sigma M_{vir}, 10^{10}M_{\sun}$   & 103     &   117    &   598    &    122    &   332   &    267    \\
 $\Sigma M_{orb}, 10^{10}M_{\sun}$   & 200     &   197    &   388    &    230    &   153   &    322    \\
 $R_0$, Mpc          &0.94$\pm$.10&   1.05$\pm$.07   &   1.26$\pm$.15   &   0.90$\pm$.10    &  0.70$\pm$.10   &   0.63$\pm$.10    \\
 $M_t , 10^{10}M_{\sun}$       & 121$\pm$38     &   169$\pm$34    &   292$\pm104$    &    106$\pm$35    &    50$\pm$22   &    37$\pm$18     \\
 $L_t , 10^{10}L_{\sun}$       & 10.1    &    6.1   &    7.9   &     5.8   &    5.6  &    2.0    \\
 $M_t /L_B$          &  12     &    28    &    37    &     18    &     9   &    19     \\
\hline
\enddata
\end{deluxetable}
\end{document}